\documentstyle[titlepage,12pt]{article}

\begin{document}
\begin{titlepage}
\title{Density Perturbations in the Brans-Dicke Theory}
\author{J. P. Baptista\thanks{{\sc E-mail: plinio@cce.ufes.br}},
J. C. Fabris\thanks{{\sc E-mail: fabris@cce.ufes.br}}
and S. V. B. Gon\c{c}alves\thanks{{\sc E-mail: sergio@cce.ufes.br}}\\
Departamento de F\'{\i}sica\\
Universidade Federal do Esp\'{\i}rito Santo\\
Goiabeiras - CEP 29060 -900\\
Vit\'oria - Esp\'{\i}rito Santo\\
Brazil}
\maketitle
\begin{abstract}
We give here the calculation of density perturbations in a 
gravitation theory with a scalar field non-minimally coupled to 
gravity, i.e., the Brans-Dicke Theory of gravitation. The purpose is 
to show the influence of this scalar field on the dynamic behaviour 
of density perturbations along the eras where the equation of 
state for the matter can be put under the form $p = \alpha\rho$, 
where $\alpha$ is a constant. We analyse the asymptotic behaviour of this 
perturbations for the cases $\alpha = 0$, $\alpha = -1$, $\alpha = 
1/3$ and $\rho = 0$. In general, we obtain a decaying and growing modes.
In the very important case of 
inflation, $\alpha = -1$, there is no density perturbation, as it is well 
known.
In the vacuum phase the perturbations on the
scalar field and the gravitational field present growing modes at the 
beginning of the expansion and decaying modes at the end of this 
phase.
In the case $\alpha = 0$ it is possible,
for some negative values of $\omega$,
to have an amplification of the perturbations with a superluminal 
expansion of the
scale factor. We can also obtain strong
growing modes for the density contrast for the case where
there is a contraction phase which can have physical interest
in some primordial era.
\end{abstract}
\end{titlepage}
\section{Introduction}
The existence of a classical scalar field in Nature has been 
considered in many theories of gravitation that present alternatives 
to General Relativity. The prototype of scalar theories is the 
Brans-Dicke Theory \cite{1},\cite{2},\cite{3},\cite{4}, whose Lagrangian 
is given by:
\begin{equation}
\label{1}
S=\frac{1}{16\pi}\int{d^{4}x}\sqrt{-g}\biggr[\phi{R}-\omega\biggr(
\frac{\phi,_{\mu}\phi,^{\mu}}{\phi}\biggl)+16\pi{\cal{L}}_{mat}
\biggl]~~.
\end{equation}\\

It can be expected that the presence of the scalar field leads to 
different predictions with respect to those we obtain in General 
Relativity. In cosmology, this can lead to far reaching consequences, 
since the standard scenario given by General Relativity presents, besides 
some spectacular success, important drawbacks such as the horizon, 
flatness and structure formation problems\cite{5}.  

However, local physics limits the value of the parameter $\omega$: it 
must be greater than 500  to account the classical tests. This result has 
reduced the interest in the Brans-Dicke Theory. This situation has 
changed recently with the proposal of the extended inflation 
\cite{6},\cite{7}; in the de Sitter phase, Brans-Dicke Theory predicts 
power-law inflation instead of exponential; in fact, following D. La and 
P. Steinhardt, at the beginning of inflation the BD solutions (for $p = 
-\rho)$ approaches the Einstein-de Sitter solution. In the second stage 
of the inflation, both the scalar field and the scale factor grow by 
power law rather than exponential. This feature prevents the so-called 
"graceful exit"problem. However, in order to work, the parameter must be 
$\omega\approx 24$, contradicting observation; nevertheless this 
constraint follows from local conditions. This drawback can be overcome 
through a generalization of the original Brans-Dicke Theory, allowing the 
parameter $\omega$ to be a function of the field $\phi$ itself.

This revival of Brans-Dicke Theory leads us to ask if the problem of 
structure formation can be modified through the introduction of scalar 
field. This question has been treated in many different situations in the 
literature\cite{8},\cite{9}, \cite{10}. Here, we propose to study the 
evolution of density fluctuation in the traditional Brans-Dicke Theory in 
the different phases of the Universe. Even if we consider $\omega$ as a 
constant, this analyses can furnish many insights in how the scalar field 
modifies the main conclusions about gravitational instability obtained 
employing General Relativity. In particular we will see that in de Sitter 
phase, the scenario differs substantially from the traditional one. In 
the other phases, however, the differences are much less important. As it 
was mentioned above, the main interest in the presence of the scalar 
field is the possibility that this field  accelerates the growth of 
density perturbations.

We will work in the Lifschitz-Khalatnikov formalism \cite{11}. The reason 
is that it furnishes essentially the same relevant results we can obtain 
employing another formalism, and also it allows for the choice of 
synchronous frames where the physical interpretation of the concerned 
quantities are easily done. We must, of course, be careful about the 
presence of unphysical modes\cite{12}. However they can be eliminated by 
performing an infinitesimal coordinate transformation. Taking care of the 
so called residual coordinate freedom, we can be sure to retain just the 
physical modes.

We will alllow for negative values for the parameter $\omega$. In fact, a 
negative
$\omega$ is what is predicted by the effective models comming from
Kaluza-Klein and Superstring theories\cite{13}.

This paper is organized as follows. In section II we present the 
background solutions of the unperturbed Universe in the following phases 
of its development: vacuum, inflation, radiation and incoherent matter. 
In section III we obtain the perturbed equations and their solutions in 
terms of Bessel functions for that phases. In section IV we calculate the 
asymptotic behaviour of the solutions for $t\rightarrow 0$ and 
$t\rightarrow\infty$. Finally, in section V we discuss the relations 
between the wavelength $\lambda$ and the particle-horizon distance $H^{-1}$.

We will assume in this article the following notations: the greek indices 
run from zero to three; the latin indices run from one to three; the 
signature is $(+,-,-,-)$; we use a Robertson-Walker metric with flat 
space section ($k=0$); the scalar field $\phi$ is a time function; the 
energy-momentum is the perfect fluid; we use the synchronous gauge which 
fix the reference frame.
\section{Background Solutions}

We assume that background Universe is spatially flat, homogeneous and 
isotropic, i. e., it is described by the Robertson-Walker metric:
\begin{equation}
\label{3}
ds^{2}=dt^{2}-a^{2}dx^{2}~~,
\end{equation}
where the $a(t)$ is a scale factor of the Universe, and $c=1$. The 
energy-momentum tensor of the background matter takes a perfect fluid form:
\begin{equation}
\label{4}
T^{\mu\nu}=(\rho+p)U^{\mu}U^{\nu}-pg^{\mu\nu}~~,
\end{equation}
with an equation of state $p = \alpha\rho$. In these expressions $\rho$ 
is a density matter, $p$ is a pressure and $\alpha$ is a constant.

From the Lagrangian density (\ref{1}), we obtain the field equations:
\begin{displaymath}
R_{\mu\nu}-\frac{1}{2}g_{\mu\nu}R=\frac{8\pi}{\phi}
T_{\mu\nu}+\frac{\omega}{\phi^{2}}\biggr(\phi;_{\mu}
\phi;_{\nu}-\frac{1}{2}g_{\mu\nu}\phi;_{\rho}
\phi;^{\rho}\biggl)+ 
\end{displaymath}
\begin{equation}
\label{180}
+\frac{1}{\phi}\biggr(\phi;_{\mu};_{\nu}-g_{\mu\nu}\Box\phi\biggl)~~,
\end{equation}
\begin{equation}
\label{170}
\Box\phi=\frac{8\pi}{3+2\omega}T~~.
\end{equation}

Inserting the metric (\ref{3}), into equations (\ref{180}) and 
(\ref{170}) we obtain the equations of motion:
\begin{equation}
\label{5}
-3\frac{\ddot{a}}{a}=\frac{8\pi}{\phi}\rho\biggr(\frac{2+\omega+3\alpha+3\alpha\omega}{3+2\omega}\biggl)+\frac{\omega}{\phi^{2}}\dot\phi^{2}+\frac{1}{\phi}\ddot\phi~~,
\end{equation}
\begin{equation}
\label{6}
\ddot\phi+3\frac{\dot{a}}{a}\dot\phi=\frac{8\pi}{3+2\omega}\rho(1-3\alpha)~~.
\end{equation}
 
The above equations must be supplemented by a conservation equation:
\begin{equation}
\label{7}
T^{\mu\nu};_{\nu}=0~~;
\end{equation}
when $\mu=0$ we have the energy conservation:
\begin{equation}
\label{8}
\dot\rho+3\frac{\dot{a}}{a}\rho(1+\alpha)=0~~,
\end{equation}
while, for $\mu=j$ we get the condition:
\begin{equation}
\label{9}
\rho,i=0~~.
\end{equation}

Background solutions can be obtained assuming that the scale factor 
$a(t)$ and the scalar field $\phi(t)$  have a power-law form:
\begin{equation}
\label{10}
a(t)\propto t^{r}~~~,~~~~~~~~\phi(t)\propto t^{s}~~.
\end{equation}

By direct substitution of the above relation in the equations (\ref{5}), 
(\ref{6}), (\ref{8}) and (\ref{9}), we have:
\begin{equation}
\label{11}
-3r(r-1)=8\pi\rho 
t^{2-s}\biggr(\frac{2+\omega(1+3\alpha)+3\alpha}{3+2\omega}\biggl)+s^{2}(1+\omega)-s~~,
\end{equation}
\begin{equation}
\label{12}
s^{2}+s(3r-1)=\frac{8\pi}{3+2\omega}\rho t^{2-s}(1-3\alpha)~~,
\end{equation}
\begin{equation}
\label{13}
\dot\rho+3\frac{r}{t}\rho(1+\alpha)=0~~,
\end{equation}
\begin{equation}
\label{14}
\rho,i=0~~.
\end{equation}

In what follows we shall consider the important  special cases: vacuum, 
inflation, radiation and incoherent matter.
\subsection{Vacuum~~~($\rho=0$)}

Here, the Universe has no ordinary matter and the energy-momentum tensor 
is null in all space-time. Our background solutions from equations 
(\ref{11}), (\ref{12}) are
\begin{equation}
\label{15}
s=1-3r~~~,~~~~~~~~r=\frac{\omega+1\pm\sqrt{1+\frac{2\omega}{3}}}{4+3\omega}~~,
\end{equation}
where $\omega>-\frac{3}{2}$.
\subsection{Inflation~~~($\alpha=-1$)}

In this scenario we have a drastic expansion of the Universe during the 
early period of the Big Bang. In this case, the background solutions are:
\begin{equation}
\label{16}
s=2~~~,~~~~~~~~r=\omega+\frac{1}{2}~~.
\end{equation}

Strictly speaking, inflation occurs for $\omega\geq \frac{1}{2}$.
\subsection{Radiation~~~($\alpha=\frac{1}{3}$)}

In this era the energy density of the Universe was tottaly dominated by 
relativistic particles. The  solutions are:
\begin{equation}
\label{18}
s=0~~~,~~~~~~~~r=\frac{1}{2}~~.
\end{equation}
As it is easily seen the above solutions are the same for the General 
Relativity by the same scenario\cite{14}.
\subsection{Incoherent Matter~~~($\alpha=0$)}

This phase is such that the Universe is dominated by nonrelativistic 
matter with negligible pressure. The solutions are:
\begin{equation}
\label{17}
s=2-3r~~~,~~~~~~~~r=\frac{2+2\omega}{4+3\omega}~~.
\end{equation}

\section{Perturbed Equations}

Our objective here is to calculate the perturbations of the equations 
(\ref{5}), (\ref{6}) and (\ref{7}), and to obtain the exact solutions for 
the perturbed equations for each scenario of the Universe. We are 
interested in the density matter perturbations so that we will only use 
the temporal components of the Einstein's equations since that component 
is directly associated with the scalar modes. Adopting the synchronous 
gauge, we put $h_{0\mu}=0$.

In order to derive the perturbed equations, we write the metric tensor as:
\begin{equation}
\label{19}
\widetilde{g}_{\mu\nu}=g_{\mu\nu}+h_{\mu\nu}~~,
\end{equation}
leading to the Ricci tensor
\begin{equation}
\label{20}
\widetilde{R}_{00}=R_{00}+\delta R_{00}~~,
\end{equation}
where 
\begin{displaymath}
\delta R_{00}=\frac{1}{a^{2}}\biggr[\ddot h_{kk}-2\frac{\dot{a}}{a}\dot 
h_{kk}+2\biggr(\frac{\dot{a}^{2}}{a^{2}}-\frac{\ddot 
a}{a}\biggl)h_{kk}\biggl]~~.
\end{displaymath}
The energy-momentum tensor is given by
\begin{equation}
\label{21}
\widetilde T^{00}=T^{00}+\delta T^{00}=T^{00}+\delta\rho~~;
\end{equation}
the energy-momentum tensor trace is
\begin{equation}
\label{22}
\widetilde T=T+\delta T=T+\delta\rho-3\delta p~~;
\end{equation}
for the  d'Alembertian of the scalar field we find
\begin{equation}
\label{23}
\delta\Box\phi=\delta\ddot\phi+a\dot ah^{kk}\dot\phi-\frac{1}{2a^{2}}\dot 
h_{kk}\dot\phi+3\frac{\dot 
a}{a}\delta\dot\phi-\frac{\bigtriangledown^{2}}{a^{2}}\delta\phi~~.
\end{equation}

Using now the following definitions\\
i) $h_{kk}=a^{2}h~~,$\\
ii) $\delta\phi=\lambda\phi~~$ with $\lambda<<1~~,$\\
iii) $\delta\rho=\Delta\rho~~$ with $\Delta<<1~~,$\\
we will get the set of perturbed equations:
\begin{equation}
\label{24}
\frac{1}{2}\ddot h+\frac{\dot a}{a}\dot 
h=\frac{8\pi}{\phi}\rho(\Delta-\lambda)\biggr(\frac{2+\omega+3\omega\alpha+3\alpha}{3+2\omega}\biggl)+\ddot\lambda+2\dot\lambda\frac{\dot\phi}{\phi}(1+\omega)~~,
\end{equation}
\begin{displaymath}
\ddot\lambda+\dot\lambda\biggr(2\frac{\dot\phi}{\phi}+3\frac{\dot 
a}{a}\biggl)+\lambda\biggr(\frac{\ddot\phi}{\phi}+3\frac{\dot 
a}{a}\frac{\dot\phi}{\phi}\biggl)-\frac{1}{2}\dot 
h\frac{\dot\phi}{\phi}-\frac{1}{a^{2}}\bigtriangledown^{2}\lambda=
\end{displaymath}
\begin{equation}
\label{25}
=\frac{8\pi}{(3+2\omega)}\frac{\Delta}{\phi}\rho(1-3\alpha)~~,
\end{equation}
\begin{equation}
\label{26}
\dot\Delta-(1+\alpha)\biggr(\frac{1}{2}\dot h-\delta U^{k},_{k}\biggl)=0~~,
\end{equation}
\begin{equation}
\label{27}
(1+\alpha)(2-3\alpha)\frac{\dot a}{a}\delta U^{j}+\delta\dot 
U^{j}(1+\alpha)=-a^{-2}\alpha\Delta_{,j}g^{jk}~~.
\end{equation}

Equations (\ref{24}), (\ref{25}) are the differential equation for the 
perturbation $h(t)$ of the gravitational field, $\lambda(t)$ is the 
scalar field perturbation and $\Delta(t)$, the perturbation for ordinary 
matter. Equations (\ref{26}) and (\ref{27}) are the perturbed equations 
derived from the conservation equation (\ref{7}). 
With the above set of differential second order equations, we will 
analyse each phase of the Universe.

In what follows, we will suppose that the perturbations have a plane 
wave  behaviour:
\begin{equation}
\label{28}
\lambda(x,t)=\lambda(t)\mbox{exp}(-i\vec{q}.\vec{x})~~\mbox{and}~~~~~~~~
h(x,t)=h(t)\mbox{exp}(-i\vec{q}.\vec{x})~~,
\end{equation}
where $q$ is the wavenumber of the perturbations.
\subsection{Vacuum~$~~(\rho=0)$}

In the Universe without matter there is only perturbations of the metric 
and scalar field. 
The two coupled linear second-order equation for $h(t)$ and $\lambda(t)$ 
are given by:
\begin{equation}
\label{29}
\frac{1}{2}\ddot{h}+\frac{r}{t}\dot{h}=\ddot\lambda+2\frac{\dot\lambda}{t}(1+\omega)(1-3r)~~,
\end{equation}
\begin{equation}
\label{30}
\frac{1}{2}\dot{h}(1-3r)=\ddot\lambda t+2\dot\lambda(1-3r)+3\dot\lambda 
r+\frac{1}{t^{2r-1}}q^{2}\lambda~~.
\end{equation}

Differentiating (\ref{30}) and removing $\dot h$ and $\ddot h$, we find 
the following linear third-order differential equation for $\lambda(t)$:
\begin{displaymath}
\stackrel{...}\lambda+\ddot\lambda\biggr[\frac{2}{t}(1+r)\biggl]+
\end{displaymath}
\begin{displaymath}
+\dot\lambda\biggr[\frac{r}{t^{2}}(16+12\omega)-\frac{r^{2}}{t^{2}}
(24+18\omega)\frac{1}{t^{2}}
(2+2\omega)+\frac{q^{2}}{t^{2r}}\biggl]+
\end{displaymath}
\begin{equation}
\label{31}
+\lambda\biggr(\frac{q^{2}}{t^{2r}}\biggl)=0~~.
\end{equation}

In order to solve the above equation, we perform the following substitution:
\begin{displaymath}
\lambda=\frac{\kappa}{t}~~,~~~~\dot\kappa=t^{r}\gamma~~,~~~~x=t^{p}~~,\mbox{with}~~p=1-r~~,
\end{displaymath}
\begin{displaymath}
\gamma=x^{m}g~~,\mbox{with}~~m=\frac{1-2r}{1-r}~~,~~~~y=\frac{xq}{1-r}~~.
\end{displaymath}
We get from equation (\ref{31}),
\begin{equation}
\label{32}
y^{2}g''+yg'+g(y^{2}-1)=0~~,
\end{equation}
i.e., Bessel equation of order 1, where the prime denotes derivatives 
with respect to the conformal time.

The solutions of (\ref{32}) are the Bessel and Neumann functions of the 
first kind:
\begin{equation}
\label{33}
g=C_{1}J_{1}(y)+C_{2}N_{1}(y)~~.
\end{equation}

The solutions for $\lambda(t)$ and $h(t)$ are given by:
\begin{equation}
\label{34}
\lambda(t)=\frac{1}{t}\biggr[\int{t^{1-r}}\biggr(
C_{1}J_{1}(y)+C_{2}N_{1}(y)\biggl)dt+C_{3}\biggl]~~,
\end{equation}
\begin{displaymath}
h(t)=\int\Biggr[\frac{1}{(1-3r)t^{2}}\biggr[6r+\frac{2q^{2}}{t^{2r-4}}\biggl]\biggr[\int{t^{1-r}}
\biggr(C_{1}J_{1}(y)+C_{2}N_{1}(y)\biggl)dt+C_{3}\biggl]+
\end{displaymath}
\begin{displaymath}
+\frac{1}{(1-3r)t^{r}}\biggr[4-6r-\frac{4}{t}(1+r)
\biggl]\biggr[C_{1}J_{1}(y)+C_{2}N_{1}(y)\biggl]+
\end{displaymath}
\begin{equation}
\label{35}
+\frac{4}{(1-3r)t^{r}}\biggr[C_{1}\dot{J}_{1}(y)+C_{2}\dot{N}_{1}(y)\biggl]\Biggl]dt+C_{4}~~,
\end{equation}
where $C_{1},C_{2},C_{3}~~\mbox{and}~~C_{4}$ are constants and $\dot 
J_{k}, \dot N_{k}$ means respectively the time derivative of Bessel and 
Neumann functions.
\subsection{Inflation~~~$(\alpha=-1)$}

The remarkable feature of this case is that the density contrast is null. 
This can be seen through equations (\ref{26})-(\ref{27}). So, we have two 
coupled linear second-order equation for $h(t)$ and $\lambda(t)$:
\begin{equation}
\label{36}
\ddot{h}+\frac{2r}{t}\dot{h}-2\ddot\lambda-\frac{4}{t}(2r+1)\dot\lambda-\frac{16\pi}{(1+r)}r\rho\lambda=0~~,
\end{equation}
\begin{equation}
\label{37}
\ddot\lambda+2\dot\lambda\biggr(\frac{2+3r}{t}\biggl)+\frac{1}{t^{2r+2}}q^{2}\lambda-\frac{1}{t}\dot{h}+\frac{16\pi}{t^{2}(1+r)}\rho\lambda=0~~.
\end{equation}

The resolution method is very similar to the vacuum case. So we find the 
following linear third-order differential equation for $\lambda(t)$:
\begin{displaymath}
\stackrel{...}{\lambda}+\ddot\lambda\biggr(\frac{5r+3}{t}\biggl)+\dot\lambda\biggr(\frac{1}{t^{2r}}q^{2}+\frac{(6r^{2}+6r-2)}{t^{2}}\biggl)+
\end{displaymath}
\begin{equation}
\label{38}
+\lambda\biggr(\frac{1}{t^{2r+1}}q^{2}+\frac{(6r^{2}-4r-2)}{t^{3}}\biggl)=0~~.
\end{equation}

By employing the transformations:
\begin{displaymath}
\lambda=\frac{\kappa}{t}~~,~~~~\dot\kappa=t^{r}\gamma~~,~~~~x=t^{p}~~~~\mbox{with}~~p=1-r~~,
\end{displaymath}
\begin{displaymath}
\gamma=x^{m}g~~~~\mbox{with}~~m=\frac{1-7r}{2(1-r)}~~,~~~~y=\frac{xq}{1-r}~~,
\end{displaymath}
we obtain the Bessel equation of the order $k$: 
\begin{equation}
\label{39}
y^{2}g''+yg'+(y^{2}-k^{2})g=0~~,
\end{equation}
where
\begin{displaymath}
k=\frac{r+3}{2(1-r)}~~.
\end{displaymath}

Its solution is
\begin{equation}
\label{40}
g(y)=C_{1}J_{k}(y)+C_{2}J_{-k}(y)~~,
\end{equation}
where $J_{k}(y)$ is the Bessel function of the order $k$.

The solutions for $\lambda(t)$ and $h(t)$ are given by:
\begin{equation}
\label{41}
\lambda(t)=\frac{1}{t}\biggr[\int{t^{\frac{1-5r}
{2}}}\biggr(C_{1}J_{k}(y)+C_{2}J_{-k}(y)\biggl)dt+C_{3}\biggl]~~,
\end{equation}
\begin{displaymath}
h(t)=\int\Biggr[t^{-\frac{1+5r}{2}}\biggr[\biggr(
\frac{1-5r}{2}\biggl)(C_{1}J_{k}(y)+C_{2}J_{-k}(y))+
\end{displaymath}
\begin{displaymath}
+t^{-\frac{5r}{2}}\biggr[\biggr(
\frac{1-5r}{2}\biggl)(C_{1}\dot{J}_{k}(y)+C_{2}\dot{J}_{-k}(y))\biggl]+
\end{displaymath}
\begin{displaymath}
+\frac{(3r+2)}{t}\biggr[-\frac{1}{t^{2}}\biggr(\int{t^{\frac{1-5r}{2}}}(C_{1}J_{k}(y)+C_{2}J_{-k}(y))dt+C_{3}\biggl)+
\end{displaymath}
\begin{displaymath}
+\frac{1}{t}\biggr[\int 
t^{\frac{1-5r}{2}}(C_{1}J_{k}(y)+C_{2}J_{-k}(y))\biggl]\biggl]+
\end{displaymath}
\begin{equation}
\label{42}
+\biggr(\frac{(6r+2)}{t^{2}}+\frac{1}{t^{2r}}q^{2}\biggl)\biggr[\int{t^{\frac{1-5r}{2}}}(C_{1}J_{k}(y)+C_{2}J_{-k}(y))dt+C_{3}\biggl]\Biggl]dt+C_{4}~~,
\end{equation}
where $C_{1},C_{2},C_{3},C_{4}$ are constants. 

In contrast with the vacuum case, $r$ is greater than 1 for 
$\omega>\frac{1}{2}$. For $r=1~(\omega=\frac{1}{2})$, equation (\ref{38}) 
becomes an Euler equation and admit the particular power-law solution:
\begin{equation}
\label{Plinio}
\lambda=t^{n},~~~~~~~n=-2\pm\sqrt{4-q^{2}}.
\end{equation}

\subsection{Radiation~~($\alpha=\frac{1}{3}$)}

By considering the background solutions for the perturbed quantities, we 
have, from equations (\ref{24})-(\ref{27})
\begin{equation}
\label{54}
\ddot{h}+\frac{1}{t}\dot{h}=2\ddot\lambda+\frac{3}{2t^{2}}(\Delta-\lambda)~~,
\end{equation}
\begin{equation}
\label{55}
\ddot\lambda+\frac{3}{2t}\dot\lambda+\frac{1}{t}q^{2}\lambda=0~~,
\end{equation}
\begin{equation}
\label{56}
\dot\Delta=\frac{2}{3}\dot{h}-\frac{4}{3}
\delta{U^{j},_{j}}~~,
\end{equation}
\begin{equation}
\label{57}
\delta U^{j}+2t\delta\dot U^{j}=-\frac{1}{2}\Delta^{,j}~~.
\end{equation}

Taking equation (\ref{55}) and performing the following transformations:
\begin{displaymath}
x=t^{p}~~,~~~~\mbox{with}~~p=\frac{1}{2}~~,
\lambda=x^{m}g~~~~\mbox{with}~~m=-\frac{1}{2}~~\mbox{and}~~~~y=2qx~~,
\end{displaymath}
we obtain the differential Bessel equation:
\begin{equation}
\label{mm}
g''+\frac{1}{y}g'+\frac{1}{y^{2}}\biggr[y^{2}-\biggr(\frac{1}{2}\biggl)^{2}\biggl]g=0~~.
\end{equation}

The solution for $\lambda(t)$ is:
\begin{equation}
\label{58}
\lambda(t)=t^{-\frac{1}{4}}\biggr(C_{1}J_{\frac{1}{2}}(y)+C_{2}J_{-\frac{1}{2}}(y)\biggl)~~,
\end{equation}
where $C_{1}$ and $C_{2}$ are constants.

Now, from equations (\ref{54}),(\ref{56}) and the equation (\ref{57}),we 
have two coupled linear differential second-order equations to solve:
\begin{equation}
\label{59}
\ddot h+\frac{1}{2t}\dot 
h-\frac{3}{2}\ddot\Delta-\frac{3}{4t}\dot\Delta-\frac{1}{2t}q^{2}\Delta=0~~,
\end{equation}
\begin{equation}
\label{60}
\ddot h+\frac{1}{t}\dot h=2\ddot\lambda+\frac{3}{2t^{2}}(\Delta-\lambda)~~.
\end{equation}

Subtracting these two equations we have:
\begin{equation}
\label{61}
\dot{h}=-3t\ddot\Delta-\frac{3}{2}\dot\Delta-\biggr(q^{2}-\frac{3}{t}\biggl)\Delta+F(\lambda)~~,
\end{equation}
where
\begin{equation}
\label{62}
F(\lambda)=4t\ddot\lambda-\frac{3}{t}\lambda~~.
\end{equation}

With the equation (\ref{61}) and equations (\ref{59}), we obtain the 
differential third-order non-homogeneous equation for $\Delta(t)$:
\begin{equation}
\label{63}
\stackrel{...}\Delta+\frac{5}{2t}\ddot\Delta+\biggr(
\frac{1}{3t}q^{2}-\frac{1}{2t^{2}}\biggl)\dot\Delta+
\biggr(\frac{1}{2t^{3}}+\frac{1}{3t^{2}}q^{2}\biggl)
\Delta={\cal{F}}(\lambda)~~,
\end{equation}
where
\begin{equation}
\label{64}
{\cal{F}}(\lambda)=\frac{1}{3t}\dot{F}(\lambda)+
\frac{1}{6t^{2}}F(\lambda)~~.
\end{equation}

Performing the transformations:
\begin{displaymath}
\Delta=\frac{\kappa}{t}~~~~\mbox{and}~~~~\dot\kappa=
t\gamma~~,
\end{displaymath}
we obtain the following differential second-order non-homogeneous equation:
\begin{equation}
\label{65}
\ddot\gamma+\frac{3}{2t}\dot\gamma+\frac{1}{3t}q^{2}\gamma={\cal{F}}(\lambda)~~,
\end{equation}
whose solution for $\Delta(t)$ is:
\newpage
\begin{displaymath}
\Delta(t)=\frac{1}{t}\Biggr[\int{t^{\frac{5}{4}}}\biggr[\bar{C_{1}}J_{\frac{1}{2}}(z)+\bar{C_{2}}J_{-\frac{1}{2}}(z)+
\end{displaymath}
\begin{displaymath}
+ 
\frac{q}{\sqrt{3}}J_{\frac{1}{2}}(z)\int{{\cal{F}}}(\lambda(t))t^{-\frac{1}{2}}\biggr(\frac{J_{-\frac{1}{2}}(z)}{W(\frac{1}{2},-\frac{1}{2})
}\biggl)dt+
\end{displaymath}
\begin{equation}
\label{66}
+\frac{q}{\sqrt{3}}J_{-\frac{1}{2}}(z)\int{{\cal{F}}}(\lambda(t))t^{-\frac{1}{2}}\biggr(
\frac{J_{\frac{1}{2}}(z)}{W(\frac{1}{2},-\frac{1}{2})}\biggl)dt
\biggl]dt\Biggl]+
\frac{1}{t}C_{3}~~,
\end{equation}
where $z=2qt^{\frac{1}{2}}/\sqrt{3}$ and 
$W(\frac{1}{2},-\frac{1}{2})=J'_{\frac{1}{2}}J_{-\frac{1}{2}}-J_{\frac{1}{2}}J'_{-\frac{1}{2}}$ the Wronskian constructed with the Bessel functions.

Equations (\ref{58}) and (\ref{66}) are the solutions for scalar field 
perturbation and for density perturbation.

\subsection{Incoherent Matter~~$(\alpha=0)$}

Here, we have the following set of coupled differential equations:
\begin{equation}
\label{43}
\ddot{h}+\frac{2r}{t}\dot{h}=\frac{16\pi}{t^{2}}(\Delta-\lambda)
\biggr(\frac{2+\omega}{3+2\omega}\biggl)+
2\ddot\lambda+4\dot\lambda\frac{(2-3r)}{t}(1+\omega)~~, 
\end{equation}
\begin{displaymath}
\ddot\lambda+\dot\lambda\biggr(\frac{4}{t}-
\frac{3r}{t}\biggl)+\lambda\biggr(\frac{2}{t^{2}}-\frac{3r}{t^{2}}
\biggl)-\frac{1}{2}\dot{h}\frac{(2-3r)}{t}+
\end{displaymath}
\begin{equation}
\label{44}
+\frac{q^{2}}{t^{2r}}\lambda=
\frac{8\pi}{t^{2}}(3+2\omega)\Delta~~,
\end{equation}
\begin{equation}
\label{45}
\dot\Delta-\biggr(\frac{1}{2}\dot{h}-\delta{U^{k}_{k}}\biggl)=0~~,
\end{equation}
\begin{equation}
\label{46}
\delta\dot U^{j}+\frac{2r}{t}\delta U^{j}=0~~.
\end{equation}

Setting the four-velocity perturbation $\delta U^{k}$ null , which, in 
this case, is allowed by one infinitesimal gauge transformation, we have:
\begin{equation}
\label{47}
(\ddot\Delta-\ddot\lambda)+\frac{2r}{t}(\dot\Delta-\dot\lambda)+\frac{(2r-2)}{t^{2}}(\Delta-\lambda)=0~~,
\end{equation}
whose solution is 
\begin{equation}
\label{48}
\Delta=\lambda+t^{2(1-r)}~~.
\end{equation}

Substituing (\ref{48}) in (\ref{44}), we obtain:
\begin{equation}
\label{49}
\ddot\lambda+\frac{2}{t}\dot\lambda+\frac{1}{t^{2r}}q^{2}\lambda=\frac{6r^{2}-13r+6}{t^{2r}}~~.
\end{equation}

Now, perfoming the transformations:
\begin{displaymath}
x=t^{p}~~\mbox{with}~p=1-r~~,~~~~\lambda=x^{m}g~~\mbox{with}~m=-\frac{1}{2(1-r)}~~, ~~~~y=\frac{xq}{1-r}~~,
\end{displaymath}
we obtain the differential second-order non-homogeneous equation:
\begin{equation}
\label{50}
y^{2}g''+yg'+\biggr[y^{2}-\biggr(\frac{1}{2(1-r)}\biggl)^{2}\biggl]g=\frac{6r^{2}-13r+6}{(1-r)^{2}}~~.
\end{equation}

Its solution is given by:
\begin{displaymath}
g(y)=\frac{6r^{2}-13r+6}{(1-r)^{2}}\Biggr[J_{\nu}(y)\int{\frac{1}{y^{2}}\biggr(\frac{J_{-\nu}(y)}{W(\nu,-\nu)}\biggl)dy}+
\end{displaymath}
\begin{equation}
\label{51}
+J_{-\nu}(y)\int{\frac{1}{y^{2}}\biggr(\frac{J_{\nu}(y)}{W(\nu,-\nu)}\biggl)
dy}\Biggl]+\bar{C_{1}}J_{\nu}(y)+\bar{C_{2}}J_{-\nu}(y)~~,
\end{equation}
where $\nu=-m$.

The solutions for $\lambda(t)$ and $\Delta(t)$ are:
\begin{displaymath}
\lambda(t)=\frac{6r^{2}-13r+6}{qt^{\frac{1}{2}}}\Biggr[
J_{\nu}(y)\int{\frac{1}{t^{2-r}}\biggr(\frac{J_{-\nu}(y)}{W(\nu,-\nu)}\biggl)dt}+
\end{displaymath}
\begin{displaymath}
+J_{-\nu}(y)\int{\frac{1}{t^{2-r}}\biggr(\frac
{J_{\nu}(y)}{W(\nu,-\nu)}
\biggl)dt}\Biggl]+
\end{displaymath}
\begin{equation}
\label{52}
+\biggl(\bar{C_{1}}J_{\nu}(y)+\bar{C_{2}}
J_{-\nu}(y)\biggr)t^{-\frac{1}{2}}~~,
\end{equation}
\begin{displaymath}
\Delta(t)=\frac{6r^{2}-13r+6}{qt^{\frac{1}{2}}}\Biggr[
J_{\nu}(y)\int{\frac{1}{t^{2-r}}\biggr(\frac{J_{-\nu}(y)}{W(\nu,-\nu)}\biggl)dt}+
\end{displaymath}
\begin{displaymath}
+J_{-\nu}(y)\int{\frac{1}{t^{2-r}}\biggr(\frac
{J_{\nu}(y)}{W(\nu,-\nu)}
\biggl)dt}\Biggl]+
\end{displaymath}
\begin{equation}
\label{53}
+(t^{-\frac{1}{2}})\biggl(\bar{C_{1}}J_{\nu}(y)+\bar{C_{2}}
J_{-\nu}(y)\biggr)+t^{2(1-r)}~~.
\end{equation}

The equations above are the solutions for $\lambda(t)$ and $\Delta(t)$ 
for $r\neq 1$.

On the other hand for $r=1$ equation (\ref{47})-(\ref{49}) admits the 
particular solution $\Delta=\lambda+C_{3}t^{-1}+Cte$, where
\begin{equation}
\label{Julio}
\lambda=t^{m},~~~~~~~~m=\frac{-1\pm\sqrt{1-4q^{2}}}{2}~~.
\end{equation}

\section{Asymptotic Behaviour}
  
In this section we analyse the solutions obtained above by calculating 
their asymptotic expressions\cite{15}. We can readly see that the 
arguments of the Bessel functions found in the last section are indeed 
the ratio beetwen the physical wavelength $\lambda_{f}=a/q$ of the 
perturbation and the particle-horizon distance $H^{-1}$. We have, then, 
$\lambda_{f}>H^{-1}$ when $t\rightarrow 0$ and $\lambda_{f}<H^{-1}$ when 
$t\rightarrow\infty$. These relations invert when $r>1$.

In what follows, for sake of simplicity, we have eliminated all gauge 
modes appearing in the solutions\cite{16}, which are represented by the 
constant $C_{3}$, and we have chosen null phases for the oscillating terms.
\subsection{Vacuum}

i) $t\rightarrow{0}$
\begin{equation}
\label{mmmm}
\lambda(t)\simeq{C_{1}t^{2(1-r)}+C_{2}}~~,
\end{equation}
\begin{displaymath}
h(t)\simeq{\frac{C_{1}}{1-3r}\biggl(2t^{1-2r}}+q^{2}t^{5-4r}-2rt^{-2r}\biggr)+
\end{displaymath}
\begin{equation}
\label{nnnn}
+\frac{C_{2}}{1-3r}\biggl(q^{2}t^{3-2r}+2t^{-1}+(4+2r)t^{-2}\biggr)~~;
\end{equation}
ii) $t\rightarrow{\infty}$
\begin{equation}
\label{oooo}
\lambda(t)\simeq{t^{\frac{r-1}{2}}[A_{1}\cos(t^{1-r})+A_{2}\sin(t^{1-r})]}~~,
\end{equation}
\begin{equation}
\label{pppp}
h(t)\simeq{t^{\frac{r-1}{2}}[A_{1}\sin(t^{1-r})+A_{2}\cos(t^{1-r})]}~~,
\end{equation}
with $A_{1}~\mbox{and}~A_{2}$ given by some functions of 
$C_{1}~\mbox{and}~C_{2}$. We also recall that
\begin{displaymath}
r=\frac{\omega+1\mp\sqrt{1+\frac{2\omega}{3}}}{4+3\omega}~~.
\end{displaymath}

\subsection{Inflation}

\hspace{1.0cm} i) for $t\rightarrow{0}$
\begin{equation}
\label{aaaa}
\lambda(t)\simeq{C_{1}t^{2(1-r)}+C_{2}t^{-(1+3r)}}~~,
\end{equation}
\begin{equation}
h(t)\simeq{\sum_{i=1}^{6}C_{i}t^{z_{i}}\biggl[\cos 
(t)\biggr]+\sum_{i=1}^{6}D_{i}t^{z_{i}}\biggl[\sin (t)\biggr]}~~;
\end{equation}
\hspace{1.0cm} ii) for $t\rightarrow{\infty}$
\begin{equation}
\label{cccc}
\lambda(t)\simeq{t^{-2(1+r)}[C_{1}\cos(t^{1-r})+C_{2}\sin(t^{1-r})}]~~,
\end{equation}
\begin{equation}
h(t)\simeq{\sum_{i=1}^{7}A_{i}t^{x_{i}}+\sum_{i=1}^{7}B_{i}t^{y_{i}}}~~,
\end{equation}
where $A_{i},B_{i},C_{i}~\mbox{and}~D_{i}$ are constants and the 
exponents are given by:\\\\
\begin{tabular}{llll}
$x_{1}=2-3r~~,$&$x_{2}=\frac{7}{2}-5r~~,$&$x_{3}=\frac{3}{2}-3r~~,$&$x_{4}=3-\frac{3}{2}r~~,$ \\
$x_{5}=5-\frac{7}{2}r~~,$&$x_{6}=\frac{7}{2}-\frac{7}{2}r~~,$&$x_{7}=\frac{11}{2}-\frac{11}{2}r~~;$ \\
$y_{1}=-1-4r~~,$&$y_{2}=-1-\frac{13}{2}r~~,$&$y_{3}=-3-\frac{9}{2}r~~,$&$y_{4}=-3-\frac{11}{2}r~~,$ \\ 
$y_{5}=1-\frac{11}{2}r~~,$&$y_{6}=-\frac{5}{2}-\frac{11}{2}r~~,$&$y_{7}=
-\frac{1}{2}-\frac{15}{2}r~~;$ \\
$z_{1}=-\frac{5}{4}-2r~~,$&$z_{2}=-\frac{1}{2}-3r~~,$&$z_{3}=-\frac{5}{2}-2r~~,$ \\
$z_{4}=-\frac{1}{2}-2r~~,$&$z_{5}=-2-2r~~,$&$z_{6}=-4r~~.$
\end{tabular}\\

The constants above are all combinations of $C_{1}~\mbox{and}~C_{2}$. 
Finally we recall that
\begin{displaymath}
r=\omega +\frac{1}{2}~~.
\end{displaymath}

\subsection{Radiation}

i) $t\rightarrow{0}$
\begin{equation}
\label{iiii}
\lambda(t)\simeq{C_{1}+C_{2}t^{\frac{1}{2}}}~~,
\end{equation}
\begin{equation}
\label{jjjj}
\Delta(t)\simeq{C_{1}t+C_{2}t^{\frac{1}{2}}}~~;
\end{equation}
ii) $t\rightarrow{\infty}$
\begin{equation}
\label{kkkk}
\lambda(t)\simeq{t^{-\frac{1}{2}}[C_{1}\sin(t^{\frac{1}{2}})+C_{2}\cos(t^{\frac{1}{2}})]}~~,
\end{equation}
\begin{equation}
\label{llll}
\Delta_{h}(t)\simeq{C_{1}'\sin(t^{\frac{1}{2}})+C_{2}'\cos(t^{\frac{1}{2}})}~~.
\end{equation} 

The non-homogeneous part of the equation (\ref{63}) gives, by a direct 
inspection, a mode that in the asymptotic regime decays with time.
\subsection{Incoherent Matter}

\hspace{0.5cm} i) $t\rightarrow{0}$
\begin{equation}
\label{eeee}
\lambda(t)\simeq{C_{1}+C_{2}t^{-1}+A(r,q)t^{r-2}}~~,
\end{equation}
\begin{equation}
\label{ffff}
\Delta(t)\simeq{\lambda(t)+t^{2(1-r)}}~~;
\end{equation}

ii) $t\rightarrow{\infty}$
\begin{equation}
\label{gggg}
\lambda(t)\simeq{t^{\frac{-2+r}{2}}[C_{1}\sin(t^{1-r})+C_{2}\cos(t^{1-r})]+At^{\frac{6r-5}{2}}\sin (t^{1-r}})~~,
\end{equation}
\begin{equation}
\label{hhhh}
\Delta(t)\simeq{\lambda(t)+t^{2(1-r)}}~~,
\end{equation}
we recall that
\begin{displaymath}
r=\frac{2+2\omega}{4+3\omega}~ ~.
\end{displaymath}

\section{Analyses of the Results}
\subsection{Vacuum Case}

In the vacuum case we verify that the exponent of the background 
solutions is always less than unity so that, for $t\rightarrow 0$, both 
$\lambda(t)$ and $h(t)$ have growing modes. On the other hand for 
$t\rightarrow\infty$, $\lambda(t)$ and $h(t)$ have only oscillating 
decaying modes. It means that perturbation of the scalar field and the 
perturbation of gravitational field whose wavelengths are greater than 
particle-horizon are growing perturbation in contrast with those having 
wavelengths smaller than the particle-horizon which are all decaying 
perturbations, as it is an expected result. This result generalizes those 
obtained in \cite{17}.
\subsection{Inflation Case}

We have in the present phase three cases: $r<1$, $r=1$ and $r>1$. (The 
case $r=1$ represents the transition between two regime in the same 
phase: inflationary expansion and ``soft expansion".)

\hspace{1.0cm} a)~$r=1$: this can represents the transition between two 
regime along the same phase: inflationary expansion and ``soft 
expansion". From equation (\ref{Plinio}) we have,

\hspace{2.0cm} i)~for $q<2$, $\lambda(t)$ has decayind modes,

\hspace{2.0cm} ii)~for $q>2$, $\lambda(t)$ has oscillating modes.

\hspace{1.0cm} b)~$r<1$: there is no inflationary inflation, in spite of 
the fact that $\alpha=-1$. Moreover, when $\omega<-1/2$, we will find an 
important growth of $\lambda(t)$, but a decaying scale factor. For 
$t\rightarrow 0$, $\lambda(t)$ has one growing mode and one decaying 
mode, while $h(t)$ has only decaying oscillating mode. For 
$t\rightarrow\infty$, $\lambda(t)$ has only decaying oscillating mode and 
$h(t)$ show growing modes.

\hspace{1.0cm} c)~$r>1$: this is the case where there is strictly 
inflation. The scalar filed grows at power $t^{2}$ and the scale factor 
undergoes huge expansion for $\omega>>1/2$. For $t\rightarrow 0$ all 
modes of $\lambda(t)$ are oscillating decaying and $h(t)$ has gentle grow 
modes for $1<r<2$ ($1/2<\omega <3/2$). For $t\rightarrow\infty$, 
$\lambda(t)$ and $h(t)$ are completly decreasing perturbations.
\subsection{Radiation Case}

In this case, $r=1/2$ and the background evolves as if the scalar field 
were constant. For $t\rightarrow 0,~~\lambda(t)$ and $\Delta(t)$ (density 
perturbation) undergoes gentle growing, as 
$t~~\mbox{and}~~t^{\frac{1}{2}}$. For $t\rightarrow\infty$ both 
perturbations show only decaying oscillating modes. In this radiation 
case the scalar perturbation gives a decaying contribuition to the 
corresponding GR results.

\subsection{Incoherent Matter Case}
Here, we have also three case $r=1,r<1, \mbox{and} r>1$. The behaviour of 
$\Delta(t)$ is the same of $\lambda(t)$ plus one term in form of power-law.

\hspace{1.0cm} a)~$r=1$: from particular solution from equation 
(\ref{49}) we see that

\hspace{2.0cm} i)~for $q<1/2$, $\lambda(t)$ has decaying modes;

\hspace{2.0cm} ii)~for $q>1/2$, $\lambda(t)$ has two oscillating modes 
with decaying amplitude.

$\Delta(t)$ has a supplementary constant term.

\hspace{1.0cm} b)~$r>1$: in this case for $t\rightarrow 0$, $\lambda(t)$ 
has oscillating modes, one of then with growing amplitude. The behaviour 
of $\Delta(t)$ shows also a pure decreasing mode. For 
$t\rightarrow\infty$, $\lambda(t)$ and $\Delta(t)$ have only decaying 
modes, except for the case $r>2$ ($-3/2<\omega -4/3$). We note that in 
this case we have also a superluminal expansion of the background.

\hspace{1.0cm} c)~$r<1$: only $\Delta(t)$ has a growing mode for 
$t\rightarrow 0$. $\lambda(t)$ has only decaying oscillating modes and 
$\Delta(t)$ is still increasing, for $t\rightarrow\infty$. In the case of 
$-4/3<\omega <-1$ we have $r<0$ and then $\Delta(t)$ is growing with high 
power of $t$. We observe that, for this case, we have a contracting 
Universe. So it can be consider of physical interest if it can be applied 
to some primordial phase.

Our results confirm those obtain in (\ref{19}), where it is pointed out 
that the BD solution of this case can not leads to a solution of the 
structure formation problem, when we consider values of $\omega$ imposed 
by local physics. 

\section{Conclusion}
We gave in this paper a general classification of perturbation in BD 
Theory corresponding to the phases of the expanding universe where the 
matter is described by equations of state $p=\alpha\rho$ with $\alpha = 
0$, $\alpha=-1$, $\alpha = 1/3$ and $\rho=0$. We assumed that the 
background solution for the scalar field and for the scale factor are 
that of "pure" power-law. We have found the exact solutions of the 
perturbed equations in the form of integrals of Bessel functions. The 
calculation here confirm, in general, the very known fact: perturbations 
at scales larger than particle horizon distance grow up, while for 
smaller scales the perturbation exhibits an oscillating behaviour. It is 
not true for incoherent matter where density perturbations can be always 
amplificated. 

In the inflationary phase we can see that the strong expansion of the 
universe erases the perturbation on the scalar field but the 
gravitational perturbation is growing for scales greater than 
particle-horizon distance. On the other hand, it is very important to 
emphasize that the perturbations
in the density of the matter are null in Brans-Dicke inflation. This 
recovers the main results from classical inflationary scenario.

The radiation phase do not give an expressive contribution with respect 
to the GR corresponding results. In the incoherent matter the situation 
is more rich. For some negative values of $\omega$ we can have a 
superluminal expansion and at the same time a significant amplification 
of perturbations.

\section*{Acknowledgements}

 We have benefited from many enlightening discussions 
with Prof. M. Novello and we also thank D.W.L. Monteiro for comments. 
S.V.B. Gon\c calves thanks {\it CAPES} (Brazil) for financial support. 
This work was partially supported by {\it CNPq} (Brazil).

\end{document}